# On Importance of Steganographic Cost For Network Steganography

Wojciech Mazurczyk[1], Steffen Wendzel[2], Ignacio Azagra Villares[3], Krzysztof Szczypiorski[1]

[1] Warsaw University of Technology, Warsaw, Poland, {wm, ksz}@tele.pw.edu.pl
[2] Fraunhofer FKIE, Bonn, Germany, steffen.wendzel@fkie.fraunhofer.de
[3] Universidad Publica de Navarra, Spain, inakiazagra@hotmail.es

**Abstract**. Network steganography encompasses the information hiding techniques that can be applied in communication network environments and that utilize hidden data carriers for this purpose. In this paper we introduce a characteristic called *steganographic cost* which is an indicator for the degradation or distortion of the carrier caused by the application of the steganographic method. Based on exemplary cases for single- and multi-method steganographic cost analyses we observe that it can be an important characteristic that allows to express hidden data carrier degradation – similarly as MSE (Mean-Square Error) or PSNR (Peak Signal-to-Noise Ratio) are utilized for digital media steganography. Steganographic cost can moreover be helpful to analyse the relationships between two or more steganographic methods applied to the same hidden data carrier.

## 1. Introduction

The main aim of *network steganography* is to hide secret data in legitimate transmissions of users without destroying the hidden data carrier used. The scope of the network steganography is limited to all information hiding techniques that: (a) can be applied in communication networks to enable hidden data exchange by creating a covert communication channel; (b) are inseparably bounded to the transmission process; (c) do not destroy the hidden data carrier. The main difference between "classic" steganography and that utilized in networks is that the first relies on fooling human senses and the latter tries to deceive network devices (intermediate network nodes or end-user ones).

It is important to emphasise that for a third party observer who is not aware of the steganographic procedure, the exchange of secret data inside the carrier (steganogram) remains hidden. This is possible because embedding of a secret data into a chosen carrier remains unnoticeable for users not involved in steganographic communication. Thus, not only the secret data are hidden inside the carriers, but because of the carriers' features, the *fact of the secret data exchange* is also concealed.

In network steganography a *carrier* is at least one network traffic flow. Typically, a carrier can be multi-dimensional, i.e. it offers many opportunities for information hiding (called *subcarriers*). And a subcarrier is defined as a "place" or a timing of "event" (e.g. a header field, padding or intended sequences of packets) in a carrier where secret information can be hidden using a steganographic technique (Fig. 1).

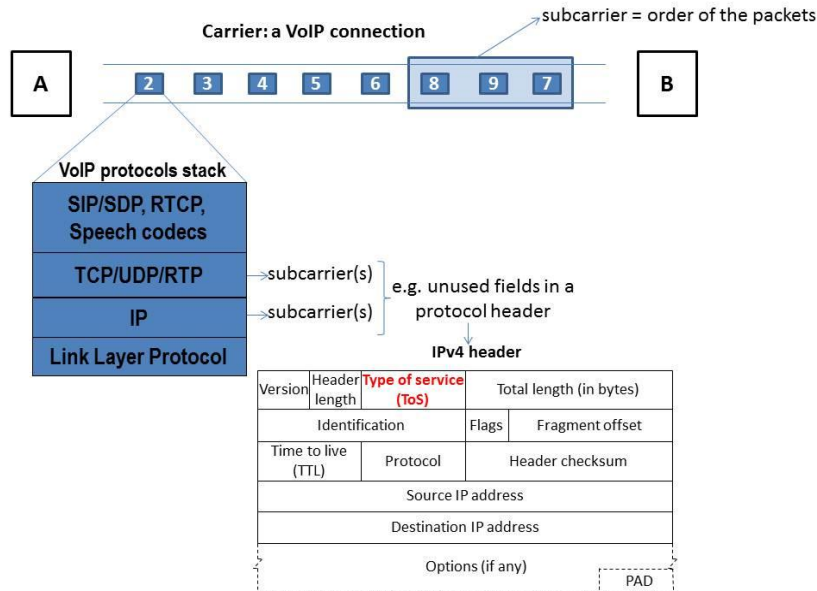

**Fig. 1** An example of carrier and subcarriers based on VoIP connection example.

The most favourable carriers for secret messages in communication networks must have two features:
- they should be popular i.e. usage of such carriers should not be considered as an anomaly itself. The more popular carriers are present and utilized in a network the better, because they mask existence of hidden communication.
- their modification related to embedding of the steganogram should not be "visible" to the third party not aware of the steganographic procedure. Contrary to typical steganographic methods, which utilize digital media (pictures, audio, and video files) as a cover for hidden data, network steganography utilizes network connections i.e. communication protocols' control elements and their basic intrinsic functionality.

Every network steganographic method can be described typically by the following set of characteristics: its *steganographic bandwidth* (also referred to as *capacity* in media steganography), its *undetectability* (also referred as *security* in literature Fridrich [1]), and its *robustness*. The term *steganographic bandwidth* refers to the amount of secret data that can be sent per unit time when using a particular method. Undetectability is defined as the inability to detect a steganogram within a certain carrier. The most popular way to detect a steganogram is to analyse the statistical properties of the captured data and compare them with the typical values for that carrier. The last characteristic is robustness that is defined as the amount of alteration a steganogram can withstand without secret data being destroyed. A good steganographic method should be as robust and hard to detect as possible while offering the highest bandwidth. However, it must be noted that there is always a fundamental trade-off among these three measures necessary (Fig. 2).

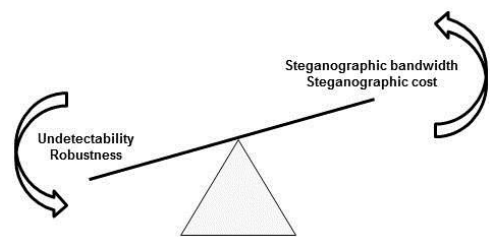

**Fig. 2** Relationship between characteristics of network steganography.

In this paper we want to emphasise that another characteristic is important while evaluating network steganography methods, namely *steganographic cost*. This characteristic indicates the degradation or distortion of the carrier caused by the application of a steganographic method. In digital media steganography, i.e. for hiding secret data in digital image, audio, or video, the parameters MSE (Mean-Square Error) or PSNR (Peak Signal-to-Noise Ratio) are utilized for this purpose. However, these parameters cannot be applied to dynamic, diverse carriers like network connections. For example, in the case of VoIP steganography methods, this cost can be expressed by providing a measure of the conversation quality degradation induced by applying a particular information hiding technique. If certain fields of the protocol header are used as the hidden data carrier, then the cost is expressed as a potential loss in that protocol's functionality. It is also possible that an information hiding method introduces steganographic cost that can be experienced in two different "planes", e.g. it introduces voice quality degradation as well as it adds additional delays to the overt traffic.

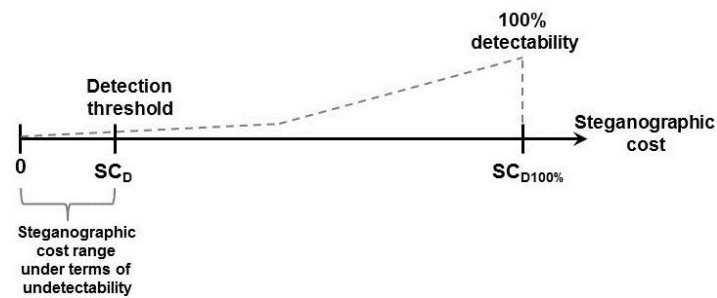

**Fig. 3** Relationship between steganographic cost and undetectability

Therefore in general we can conclude that steganographic cost affects detectability and may be responsible for loss of carrier's functionality or loss of carrier's performance (e.g. it results in longer connection or increased resources usage). The relationship between steganographic cost and detectability is explained in Fig. 3. One can imagine a steganographic cost as a "zip" as it provides a view on how exactly the carrier was affected by applying steganographic method. On the other hand detectability can be imagined as a "switch". For the certain steganalysis method when the certain level of steganographic cost is exceeded ($SC_D$) then the steganographic method becomes detectable with probability greater than 50% ("flip a coin" chance of detection) up to the point where the detection is trivial ($SC_{D100\%}$).

The effects of applying steganographic methods (steganographic cost) are threefold and form a vector for each steganographic method (Fig. 4). Some steganographic methods affect the *detectability* while others affect the *feature spectrum (e.g. reduce capabilities of the carrier, such as commands represented by header bits which are utilized by the steganographic method)* or the *performance* of the carrier; others affect multiple aspects simultaneously but to a different extent. Besides splitting performance and feature cost, both could be combined to *functionality* cost in order to achieve a twofold view on steganographic cost.

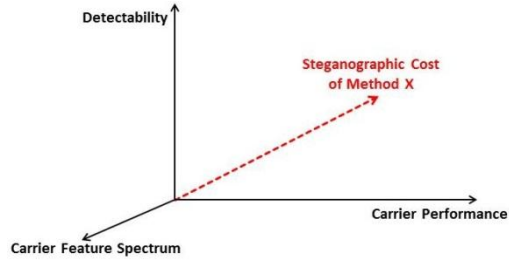
**Fig. 4** Effect of the steganographic cost

Steganographic cost can also be treated as an extension to the concept of *Minimal Requisite Fidelity* (MRF) which is introduced by Fisk *et al*. [16] in the context of active wardens. MRF is a measure of distortion which can be introduced to a potential steganographic carrier in order to *counter* a covert communication while still providing legitimate end-user acceptance of the communication. While Fisk *et al.* focus on the optimal development of *countermeasures*, our work is the first to discuss carrier degradation or distortion from the steganographer's perspective.

Another point where observation of steganographic cost can be important is when more than one method is applied to the same carrier. In this case steganographic cost allows observing the relationships between steganographic methods applied to the same hidden data carrier.

Our **main contributions** in this paper are two-fold:
- Firstly, it is a detailed analysis of the steganographic cost influence on the data carrier and its role for network steganography on the example of experimental results of existing methods: LACK (Lost Audio Packets Steganography) [2] and RSTEG (Retransmission Steganography) [3]. As in the current state-of-art for network steganography there is no characteristic defined that allows to indicate the degradation or distortion of the carrier caused by the application of the steganographic method (similarly like MSE or PSNR is used for digital media steganography) that is why we propose steganographic cost to fill that gap. In that sense our work can be also treated as a complement of Fisk's concept of *Minimal Requisite Fidelity*.
- Secondly, we propose to utilize steganographic cost to analyse the relationships between two or more steganographic methods applied to the same carrier. For this case despite general considerations we show an interesting situation which is called *super-position steganography* (and was originally introduced in [17]) where at least two methods applied simultaneously to the same carrier affect each other in such a way that the resulting total cost is lower than the overall cost of these two methods when applied alone. We illustrate this by presenting original experimental results.

To authors' best knowledge this is a first approach that deals with these two matters on the general level.

The rest of the paper is structured as follows. Section 2 focuses on the analysis of examples of single-method steganographic cost, while in Section 3 examples of multi-method steganographic cost are given. Section 4 presents experimental results for two multi-method steganographic cost scenarios. Finally, Section 5 concludes our work.

## 2. Single-method steganographic cost analysis

### 2.1 Lost Audio Packets Steganography (LACK)

LACK is an IP telephony steganographic method that was originally proposed in [9] and is currently considered as a state-of-the-art VoIP steganography technique [10]. It operates by modifying both, RTP packets from the voice stream as well as their time dependencies. It takes advantage of the fact that in typical multimedia communication protocols, like RTP, excessively delayed packets are not used for the reconstruction of transmitted data at the receiver i.e. the packets are considered useless and are discarded.

The overview of LACK's operation is presented in Fig. 5: at the transmitter (Alice), one RTP packet is selected from the voice stream and its payload is substituted with the secret message – the steganogram (1). Then, the selected audio packet is intentionally delayed prior to its transmission (2). Whenever an excessively delayed packet reaches a receiver unaware of the steganographic procedure, it is discarded. If the receiver (Bob) is aware of the hidden communication, instead of dropping the received RTP packet, it extracts the payload (3).

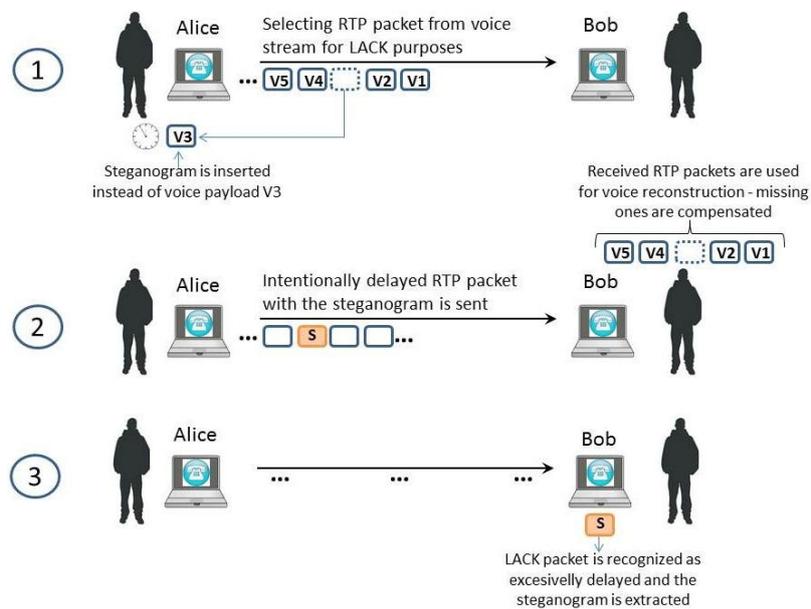

**Fig. 5** The idea of LACK

From the way LACK operates it can be deduced that its steganographic cost can be expressed as an *increased level of packet loss* (due to utilization of some of the packets for steganographic purposes) and in result it causes *decreased voice quality*. It is obvious that if too many RTP packets are selected for steganographic purposes then the resulting voice quality will be degenerated to such an extent that the steganalysis will be trivial to perform. However, in cases where care is taken while selecting the RTP packets for LACK purposes ($p_{LACK}$) the elevation of the overall packet loss level can be controlled and, for example, adjusted to the network conditions ($p_N$) to not to reach the defined detection threshold $p_T$ (Fig. 6). This will result in much smaller voice quality distortions making steganalysis significantly more demanding.

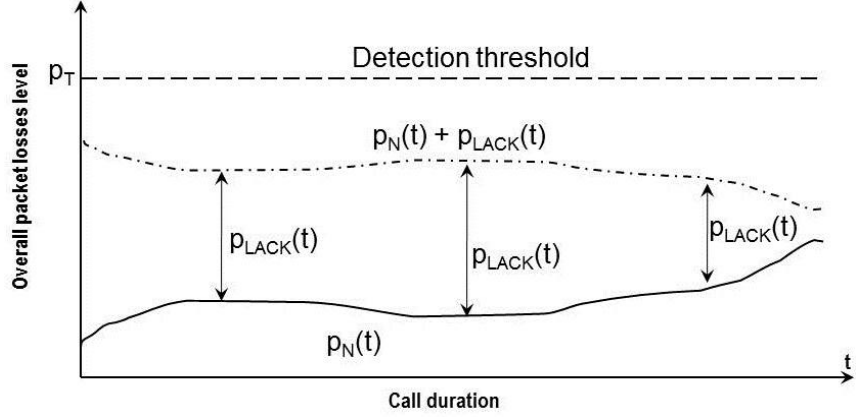

**Fig. 6** The impact of LACK on the total packet loss probability

Therefore we can express LACK's steganographic cost ($SC_{T-LACK}$) as $\Delta MOS$ which is a drop in voice quality expressed in MOS (Mean Opinion Score) scale [14] as a difference in quality of the voice signal (RQ) without and with LACK applied (LQ):

$$SC_{T-LACK}(t) = \Delta MOS(t) = RQ(t) - LQ(t)$$

Table I provides a summary and characteristics of the voice codecs used in experiments for LACK in [2]. A variety of codecs were chosen to provide a comparative analysis of possible IP telephony call configurations – the choice involved selection of voice codec and appropriate data rate (from 8 to 64 kbits). The voice codecs used in the experiment were: G.711 A-law [12], GSM-FR (Full Rate) [11] and Speex (8 and 24.6 kbits) [13] (other details on test-bed and methodology can be found in [2]).

Table I Speech codecs used in the experimental evaluation [2]

| Voice codecs | G.711 A-law | Speex I | GSM-FR | Speex II |
|---|---|---|---|---|
| Bit rate [kbit/s] | 64 | 24.6 | 13.2 | 8 |
| RTP packet every [ms] | 20 | 20 | 20 | 20 |
| Voice payload size [bytes] | 160 | 61.5 | 33 | 20 |

Based on the experimental evaluation from [2] the resulting LACK's steganographic cost for different popular VoIP codecs is presented in Fig. 7.

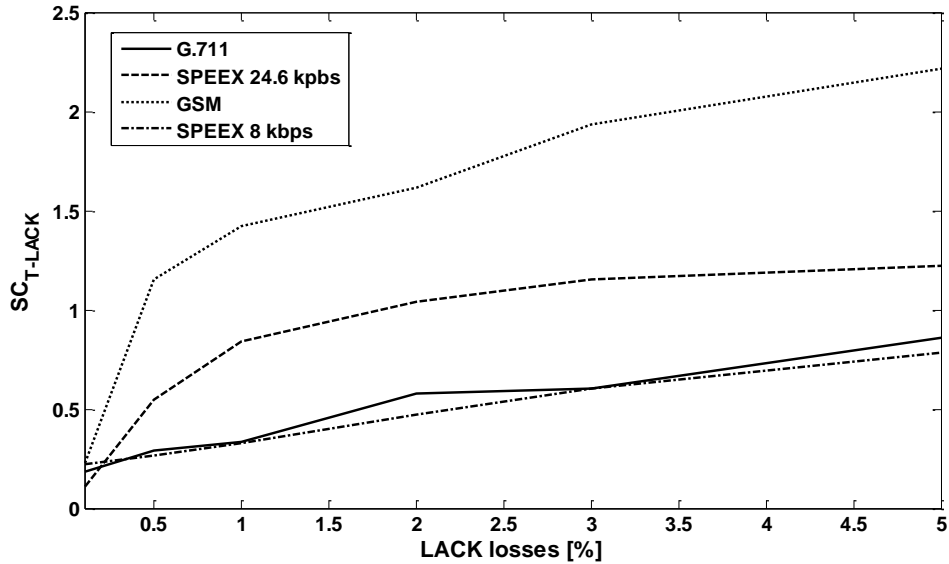

**Fig. 7** Steganographic cost for LACK for different VoIP codecs

From the results presented above we can observe that depending on how the detection threshold will be set and which voice codec will be chosen the resulting steganographic cost can be analysed in two ranges: before and after the detection threshold. For example, if $SC_{T-LACK}= \Delta MOS=0.5$ then for GSM and Speex 24.6 kbps codecs we can conclude that the introduced packets losses must be kept to minimum to avoid detection (<0.3%). However, for the other two codecs Speex 8 kbps and G.711 LACK utilization will be undetectable until intentionally introduced losses do not exceed 2-2.5%. Therefore by inspecting the steganographic cost caused by LACK from the beginning of its application we can identify the degradation of the hidden data carrier even when it is below the detection threshold.

## 2.2 Retransmission Steganography (RSTEG)

RSTEG (Retransmission Steganography) [3] is a steganographic method that is intended for a broad class of protocols that utilise retransmission mechanisms. The main innovation of RSTEG is to not acknowledge a successfully received packet in order to intentionally invoke retransmission. The retransmitted packet of user data then carries a steganogram in the payload field.

The overview of RSTEG functioning for retransmission mechanism based on timeouts is presented in Fig. 8. It is worth noting that RSTEG can also be successfully applied to other retransmission mechanisms in TCP, such as FR/R (Fast Retransmit and Recovery) or SACK (Selective Acknowledgement).

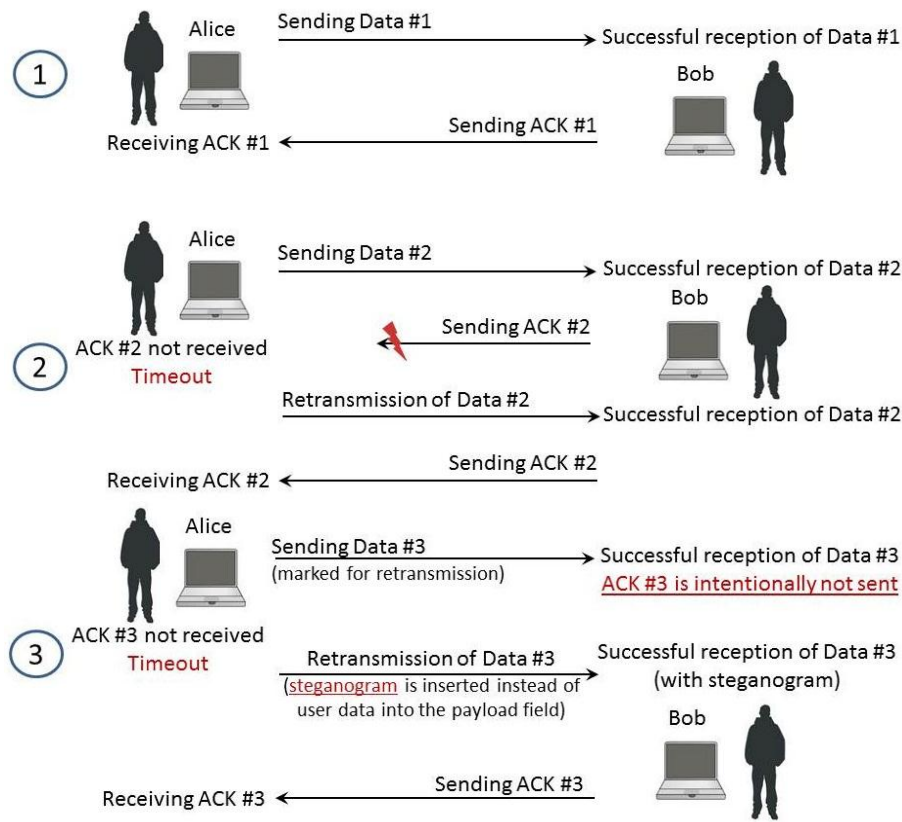

**Fig. 8** Generic retransmission mechanism based on timeouts (1, 2); RSTEG (3)

In a simplified situation, a typical protocol that uses a retransmission mechanism based on timeouts obligates a receiver to acknowledge each received packet. When the packet is not successfully received, no acknowledgment is sent after the timeout expires and so the packet is retransmitted (Fig. 8, case 1-2).

RSTEG is based on a retransmission mechanism to exchange secret data. If both sides of communication are aware of the steganographic procedure then they reliably exchange packets during their connection e.g. they transfer a file. At some point during the connection after successfully receiving a packet, the receiver intentionally does not issue an acknowledgment message. In a normal situation, a sender is obligated to retransmit the lost packet when the timeframe within which packet acknowledgement should have been received expires. In RSTEG, a sender replaces an original payload with a secret data instead of sending the same packet again. When the retransmitted packet reaches the receiver, he/she can extract the hidden information (Fig. 8, case 3).

From the description of RSTEG functioning provided above it can be deduced that its steganographic cost can be expressed as an increased level of retransmissions. Retransmissions in IP networks are a "natural phenomenon", and so intentional retransmissions introduced by RSTEG are challenging to detect if they are kept at a reasonable level.

Therefore we can express RSTEG's steganographic cost ($SC_{T\text{-}RSTEG}$) as

$$SC_{T\text{-}RSTEG} = R_D = R_{N\text{-}RSTEG} - R_N,$$

where $R_D$ (Retransmissions Difference) denotes the difference between retransmissions in the network after applying RSTEG ($R_{N\text{-}RSTEG}$) and in the network before applying RSTEG ($R_N$).

Based on experimental results in [15] that were achieved using a proof-of-concept TCP-based RSTEG implementation (assuming that network retransmission probability equals 3%, other details on test-bed and methodology can be found in [15]) the resulting $R_D$ is presented in Fig. 9 (the shape of the is explained in [15] and it is not important for our consideration thus it is omitted).

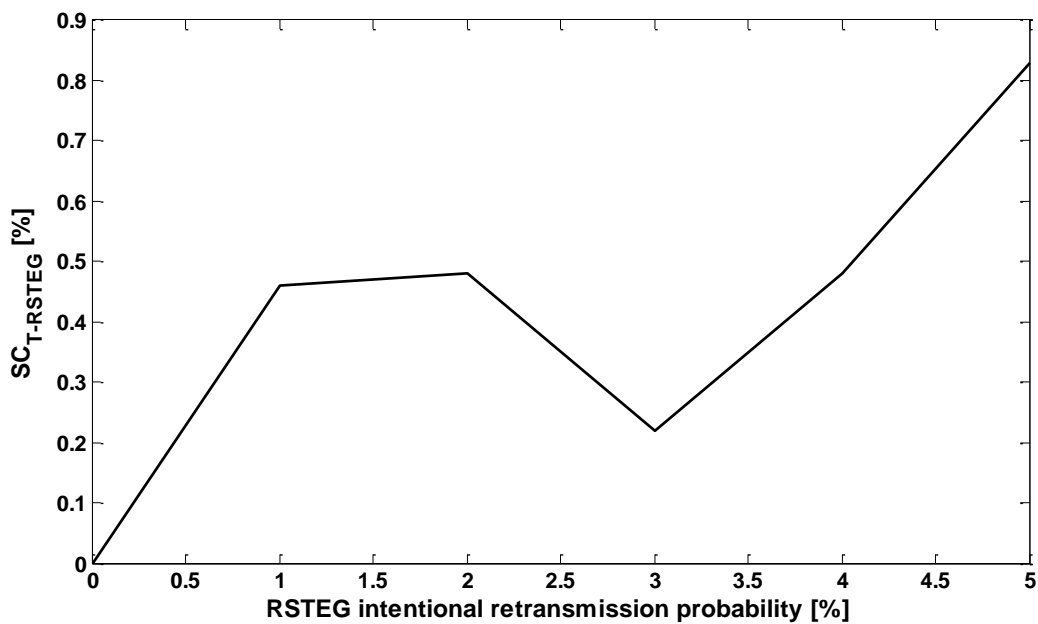

**Fig. 9** Steganographic cost for TCP-based RSTEG

Similar conclusions can be drawn for RSTEG like those presented in Sec. 2.1 for LACK. It is possible to observe the degradation of the hidden data carrier right from the beginning of the utilization of the RSTEG (Fig. 9). The steganographer is able to assess the level of the intentional retransmission for a given detection threshold in order to avoid detection and keep the steganographic cost at a reasonable level.

## 3. Multi-method steganographic cost analysis

Section 2 covered the scenario of applying one steganographic method to one hidden data carrier. Let us discuss the extension of our scenario by applying multiple methods simultaneously. For the sake of simplicity we assume that two different methods are utilized on the same hidden data carrier. The same observation that can be made for two methods can be extended to *n* methods applied to the same carrier (of course *n* is not a high number as it is difficult to design many methods that simultaneously applied will not destroy a carrier completely).

In general, two cases are possible:
- (C1) Both methods are applied to the same subcarrier,
- (C2) Both methods are applied to the different subcarriers.

It is worth noting that it is also possible that in both cases one method depends on the other, i.e. overt traffic modified by one of the methods is treated as a carrier for the other method. Such relationships between two or more methods applied to the same hidden data carrier were researched by Frączek *et al.* and are referred to as MLS (Multi-Level Steganography) [2].

In general, in case C1 the subcarrier tends to degenerate faster when both methods influence it by introducing their steganographic costs in comparison with case C2. It is due to the accumulation of the steganographic cost that influences the same subcarrier. For this case the overall steganographic cost ($SC_T$) can be expressed as

$$SC_T = SC_{S1-1} + SC_{S1-2}$$

where $SC_{S1-1}$ denotes steganographic cost of the first method applied to subcarrier S1 and $SC_{S1-2}$ denotes steganographic cost of the second method applied to the same subcarrier.

So if we consider *n* steganographic methods applied to the same subcarrier then

$$SC_{T(C1)}(n) = \sum_{n=1}^{n} SC_{S1-n}$$

Of course when utilized subcarriers are different (case C2) the steganographic cost of each steganographic method applied can be express in different units. For example, let us consider a VoIP connection: if we apply one steganographic method that affects voice quality and the second that utilizes some of the protocol header's fields then the overall steganographic cost will form a vector of steganographic cost as obviously they cannot be simply added. Therefore the steganographic cost can be expressed in this case as follows:

$$SC_{T(C2)}(n) = \begin{bmatrix} SC_{S1-1} \\ SC_{S1-2} \\ ... \\ SC_{S1-n} \end{bmatrix}$$

However, the more steganographic methods are applied even to different subcarriers the bigger chance for successful detection if various aspects of the hidden data carrier are subject to steganalysis.

In general, the following conditions must be given in case of (C1) to decrease the steganographic cost using two methods simultaneously:
1. two hiding methods utilize the same subcarrier or related subcarriers,
2. the hiding methods utilize the subcarrier(s) in a different way,

3. the hiding methods do not collide i.e. they do not negatively affect the other method and thus in result increase the steganographic bandwidth,
4. hiding method 2 modifies the embedding area of method 1 in a way that it benefits from the caused steganographic cost of method 1 in a way that less additional steganographic cost is created as if method 2 would be applied to the subcarrier without method 1.

In the remainder we concentrate on multi-method steganographic cost analyses for the case C1 to show how steganographic costs of different methods applied to the same subcarrier can interact.

**3.1 IP Header Fragmentation-based example**

Let us consider an example case where two simple steganographic methods F1 and F3 as defined in [5] are applied to an IP-based traffic flow:

Method F1 relies on the parity of the number of fragments that the packet was divided into. SS (Steganogram Sender) is the source of the fragmentation and controls this process. SS inserts a single bit of hidden data by dividing each of IP packets into the predefined number of fragments. For example, if the number of fragments is even then it means that a binary "0" is transmitted, otherwise a binary "1" (Fig. 10). The hidden data extraction is obvious as after the fragment's reception SR utilizes the number of the fragments of each received IP packet to determine the hidden data.

Of course if the statistical steganalysis based on number of fragments is performed to detect irregularities in the number of each packet's fragments the F1 method is not hard to detect.

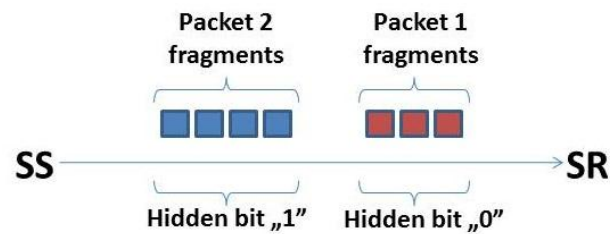

**Fig. 10** F1 steganographic method example

The second method (F3) utilizes legitimate fragments with steganogram inserted into payload for higher steganographic bandwidth and harder detection. SS is the source of the fragmentation and controls the process. During the fragmentation SS inserts secret data instead of inserting user data into the payload of selected fragments.

To make the steganographic fragments distinguishable from others yet hard to detect the following procedure was introduced. If SS and SR share a secret Steg-Key (SK) then for each fragment chosen for steganographic communication the following hash function (H) is used to calculate the Identifying Sequence (IS):

$$IS = H(SK || FragmentOffset || Identification)$$

where *Fragment Offset* and *Identification* denote values from these IP fragment header fields and || bits concatenation function. For every fragment used for hidden communication the resulting IS will have a different value due to the values change in a Fragment Offset field. All IS bits or only selected ones are distributed across the payload in a predefined manner. Thus, for each fragment SR can calculate the appropriate IS and verifies if it contains secret or user data. If the verification is successful then the rest of the payload is

considered as hidden data and extracted. Then SR does not utilize this fragment in the reassembly process of the original IP packet.

Fig. 11 illustrates an example of the proposed steganographic method. The IP packet with ID 345 is divided into four fragments (FR1-FR4). Fragment FR2 is used for steganographic purposes, so inside its payload secret data is inserted together with the correct IS. Values in the Fragment Offset and Identification fields remain the same as in other legitimate fragments. While reassembling the original packet, SR merges payloads P1, P2 and P3, omits fragment F2, and uses it only to extract secret data.

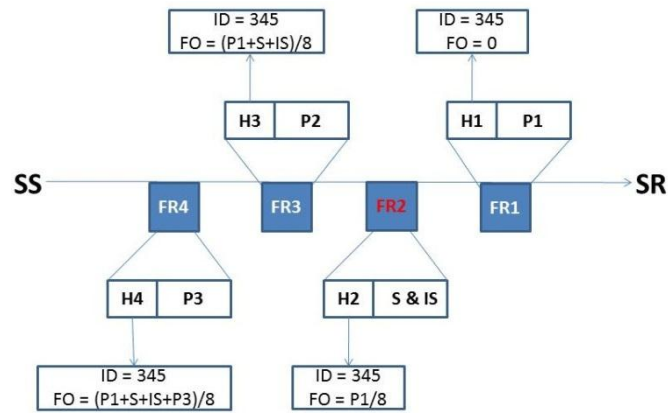

**Fig. 11** F3 steganographic method example (H – header, P – payload, S – Secret data)

Now let us consider the case when the combined F1 and F3 methods are applied simultaneously to the same hidden data carrier i.e. the same IP-based traffic flow (Fig. 12). Because F1 modulates a number of fragments that the packet is divided into and F3 inserts fake fragments then the total steganographic cost will decrease and detectability decrease, too.

When two methods applied simultaneously to the same carrier result in an overall steganographic cost decrease we refer to it as *super-position steganography*.

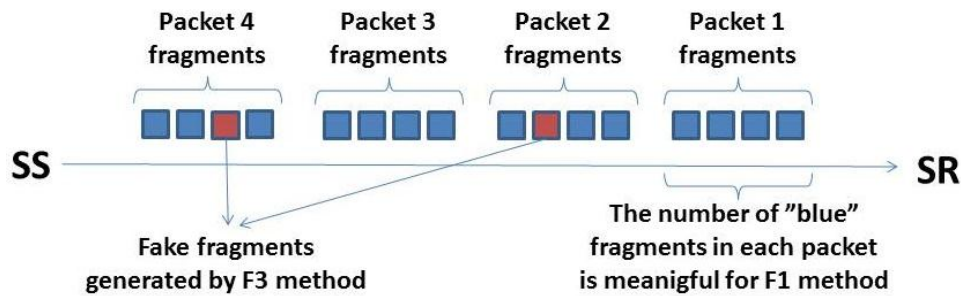

**Fig. 12** Simultaneous utilization of F1 and F3 methods

### 3.2 IPv4/IPv6 options/extensions headers-based example

Again, we consider two methods F4 and F5. F4 embeds hidden information into two IPv4 options or into IPv6 extension headers. The embedding of hidden information in IPv4 options was shown in [6] and the placement of hidden data into the IPv6 destination option was presented in [7]. Method F5 encodes hidden information by manipulating the IPv4 option's or IPv6 extension header's order in a packet like presented in [8] for the DHCP options.

Both methods, F4 and F5, can be combined to operate simultaneously utilizing the same subcarrier i.e. the IPv4 options or the IPv6 extension headers (Fig 13). If the order of the options or of the extension headers is not of relevance (e.g. because they are not interpreted and thus skipped by the hosts), the steganographic cost caused by F4 is not increased if F5 is applied as well as only the order of the non-interpreted options is changed. Therefore, no additional distortion of the carrier is caused and the total steganographic cost per steganographic method is decreased.

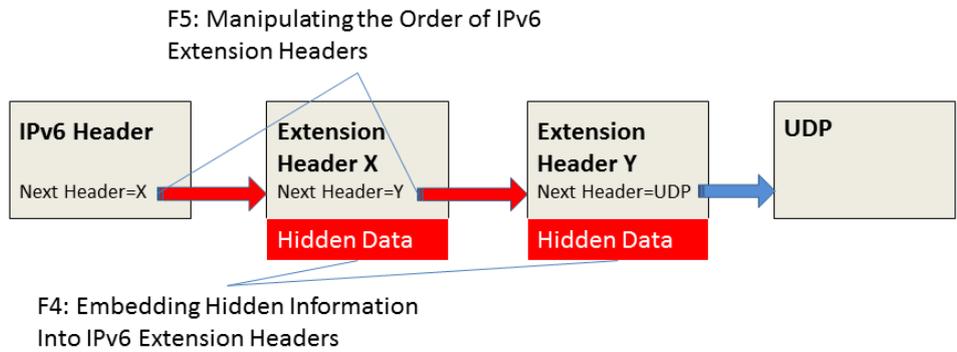

**Fig. 13** Simultaneous utilization of F4 and F5

### 3.3 Plaintext header triple method example

In order to show the feasibility of simultaneously combining more than two methods, we give the example of a HTTP request header. Method F6 therefore changes the case of the header fields, F7 changes the order of header fields, and F8 changes the number of header fields (Fig. 14). Only method F8 introduces steganographic cost by increasing the header's size and thus the available space for the remaining payload. Methods F6 and F8 modify the already created header elements of F8 without degrading the functionality of the protocol or the performance of the request. Therefore, methods F6 and F7 add no additional steganographic cost, or are, in other words *zero cost methods* in combination with F8

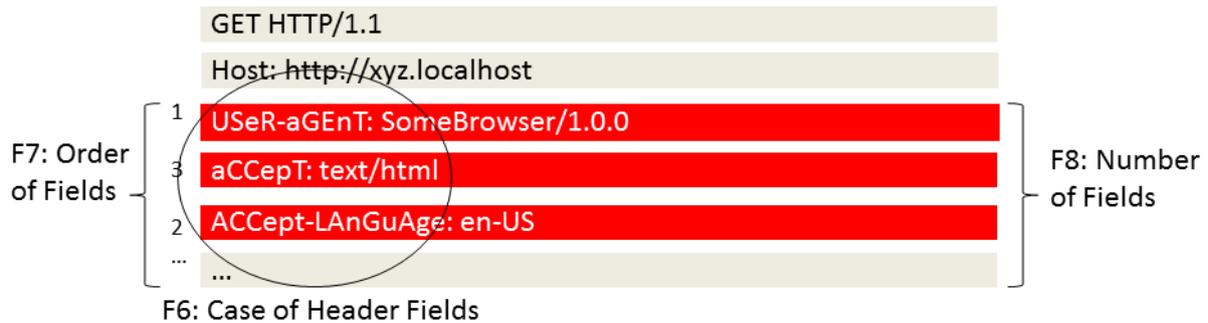

**Fig. 14** Three steganographic methods applied to the HTTP request header

### 4. Multi-method steganographic cost experimental results

Our multi-method concept was evaluated in a test-bed comprising two machines in a controlled LAN (Local Area Network) environment. The following two scenarios were realized:
- A. The IP packet fragmentation-based steganography scenario (described in Sec. 3.1) where four cases were considered in which fragmentation was performed on the stream of packets with:
  - No steganographic method applied (C1),

- Only with F1 method applied (C2),
- Only with F3 method applied (C3),
- Both F1 and F3 methods applied – super-position steganography example (C4).

B. The HTTP header-based steganography scenario (described in Sec. 3.3) where five cases were considered:
- No steganographic method applied (C5).
- Only with F6 method applied (C6),
- Only with F7 method applied (C7),
- Only with F8 method applied (C8),
- All three methods F6-8 applied (C9).

For each scenario, an own proof-of-concept implementation of the steganographic methods described in Sec. 3.1 and 3.3 was carried out. Measurements for each case in each scenario were repeated 10 times and only average values are presented. It must be emphasized that some of the implemented methods are not characterized by high undetectability but they were chosen to easily illustrate the concept of steganographic cost.

Depending on the scenario the following measures that represent steganographic cost were captured:
- *Total connection time* for both implemented scenarios,
- distribution of the *fragment sizes* for Scenario A,
- distribution of the *HTTP header sizes* for Scenario B.

In the Scenario A, for every case a total number of 2.400 packets was subject to fragmentation and transmitted. Each fragment that was created was 500 bytes to avoid trivial detection. If a steganographic method adds a fake fragment then its payload is marked using *IS* (cf. Sec. 3.1).

Table II Experimental results for Scenario A

|  | C1 | C2 | C3 | C4 |
|---|---|---|---|---|
| Connection time | 74.78 | 73.75 | 80.77 | 78.15 |
| Standard deviation for connection time | 0.79 | 0.63 | 0.59 | 0.37 |
| Total number of fragments | 7.200 | 8.498 | 9.600 | 9.600 |
| Number of fragments per packet | 3 fragments for all packets | 4 fragments for 1.498 packets 3 fragments for 902 packets | 4 fragments for all packets | 4 fragments for all packets |

By analysing the obtained experimental results (Table II) we can observe that the total connection time for the case when method F1 is applied is similar to the case when no steganography is applied. However, when we apply method F3 the connection time lasts about 5 seconds longer. It must be noted that when both methods are applied (super-position steganography example) the intuition is that the connection will last even longer. However, the resulting connection time is only about 3 seconds longer. Therefore the duration of the connection in case of the joint methods is shorter as in case of F3 applied alone. This implies that if methods F1 and F3 are combined the resulting steganographic cost is lower as compared to the steganographic cost of the single method (the one that introduces higher steganographic cost). This is the effect we call super-position steganography.

When we compare the distribution of the number of fragments per packet the situation is similar. The F1 method (case C2) introduces irregularities in the number of fragments per packet, while F3 increases the overall number of fragments per packet. Since the third-party observer does not possess the knowledge of how many fragments the packets will be divided into in advance, the F3 technique can be considered less detectable. However it must be noted that in the joint-method's case the resulting number of fragments per packet is the same as for the case when F3 is applied alone. It is the same number of fragments since irregularities introduced by F1 are "smoothed" by the second method making the overall steganographic cost for C4 the same as for C3. Therefore the overall steganographic cost is not elevated.

In the Scenario B for every case the total of 900 packets were transmitted and modified using F6-F8 methods. The obtained experimental results are presented in Table III.

Table III Experimental results for Scenario B

|  | C5 | C6 | C7 | C8 | C9 |
|---|---|---|---|---|---|
| Connection time | 67.34 | 73.85 | 73.62 | 73.18 | 73.57 |
| Standard deviation for connection time | 0.73 | 0.72 | 0.49 | 0.62 | 0.6 |
| HTTP header size | 178 B for all packets | 178 B for all packets | 178 B for all packets | 178 B for 497 packets 154 B for 403 packets | 178 B for 511 packets 154 B for 389 packets |

By inspecting the overall connection time we can observe that after applying each of the steganographic methods alone the resulting connection time increases by about 6 seconds. The same result is achieved for the combined methods case (C9). Therefore simultaneous utilization of all three methods does not influence the total connection time. That is why such a situation is called *zero cost steganography*, as adding additional methods to the existing one does not influence the resulting steganographic cost.

F8 operates by introducing irregularities in HTTP headers sizes. For C9, where two more methods are added, the irregularities are still present but they are similar as in case of F8 applied alone. Thus we can conclude that in this case the resulting total steganographic cost is not higher than in case of method F8 applied alone.

## 5. Conclusions

We introduce a metric called steganographic cost describing the degradation or distortion of a carrier by one or more steganographic methods. We propose to take the evaluation of steganographic cost into account when a steganographic method is evaluated – in addition to the traditional measures steganographic bandwidth, detectability, and robustness. Our work complements the existing approach on Minimal Requisite Fidelity (MRF) that introduced a means to describe the distortion of a carrier by an active warden.

Our obtained experimental results show that it is feasible to combine multiple steganographic methods to the same carrier in a way that the overall steganographic cost caused by these methods is lower as in case of a separate combination of these methods (super-position steganography). Results additionally show that multiple steganographic methods can be combined with another method without causing any additional cost, which is a special case of super-position steganography called zero cost steganography.